\documentclass[twocolumn,pra,aps,showpacs]{revtex4}
\usepackage{amsfonts}
\usepackage{amsmath}
\usepackage{epsfig}

\begin{document}

\title{Classical Trajectory Perspective on Double Ionization Dynamics of
Diatomic Molecules Irradiated by Ultrashort Intense Laser Pulses}
\author{ Di-Fa Ye$^{1,2}$}
\author{ Jing Chen$^{1}$ }
\author{ Jie Liu$^{1,}$ }
\email[Email: ]{liu_jie@iapcm.ac.cn}
\affiliation{1.Institute of
Applied Physics and Computational Mathematics, Beijing
100088, P. R. China\\
2.Graduate School, China Academy of Engineering Physics, Beijing 100088, P.
R. China}

\begin{abstract}
In the present paper, we develop a semiclassical quasi-static model
accounting for molecular double ionization in an intense laser
pulse. With this  model, we achieve insight into the dynamics of two
highly-correlated valence electrons under the combined influence of
a two-center Coulomb potential and an intense laser field, and
reveal the significant influence of molecular alignment on the ratio
of double over single ion yield. Analysis on the classical
trajectories unveils sub-cycle dynamics of the molecular double
ionization. Many interesting features, such as the accumulation of
emitted electrons in the first and third quadrants of parallel
momentum plane, the regular pattern of correlated momentum with
respect to the time delay between closest collision and ionization
moment, are  revealed and successfully explained by back analyzing
the classical trajectories. Quantitative agreement with experimental
data over a wide range of laser intensities from tunneling to
over-the-barrier regime is presented.
\end{abstract}

\maketitle

\section{Introduction}

Atoms and molecules exposed to high-intensity ultrashort laser
pulses have been  attracting much attention during the past ten
years and plenty of striking phenomena  emerged such as multiphoton
ionization (MPI), above threshold ionization (ATI), high order
harmonic generation (HHG) and nonsequential double ionization (NSDI)
\cite{review,book}. The dynamics behind the  NSDI is most
complicated because it deals with two highly entangled valence
electrons \cite{nature}. In contrary to the situation of single
ionization where the experimental data can be accurately predicted
by the single active electron (SAE) model \cite{landau}%
, the NSDI data are usually  higher  by many orders of magnitude
than the sequential Ammosov-Delone-Krainov (ADK) tunneling theory
that bases on the SAE approximation. This fact suggests that the
electron-electron correlation plays a crucial role in NSDI physics.
Diverse correlation mechanisms like the shake-off \cite{shakeoff}
and the recollision \cite{corkum,coulumb} were proposed to account
for the electron correlation in the experimental observations. It is
commonly believed now that the rescattering is the dominant
mechanism for the excessive  atomic double ionization (DI)
yields\cite{rescatter}.

Compared to atomic case, the NSDI in the molecules is more
complicated because of the additional freedom that molecules
have\cite{guo,cornaggia}. In addition to the typical strong-field
phenomena like excessive DI yields in the tunneling regime, high
harmonic generation of the driving laser field, and momentum
correlation between the two emitted electrons, some new phenomena
such as bond softening \cite{soft}, zero photon dissociation
\cite{dissociate}, and alignment dependence of DI yield
\cite{alnaser,eremina,zeidler} were observed in recent molecular
experiments.

However, these experimental data for molecular double ionization is
far from well understood in theory. The complex dynamics of
correlated $e-e$ pair responding to two-center nuclear attraction
and laser force poses a great challenge to any quantum theoretical
treatment. For instance, a fully-dimensional quantum-mechanical
computation from first principle is very very time expensive even
for the simpler case of highly symmetric atoms \cite{taylor}. This
leaves approximate approaches developed recently, such as
one-dimensional quantum model \cite{pegarkov}, many-body S-matrix
\cite{beck} and simplified classical methods \cite{saddlepoint}. In
either case, the complex electron dynamics which is crucial for
molecular DI is not fully explored and the theoretical results can
not account for experimental data quantitatively.

In this paper, we explore a feasible semiclassical theory, following
the treatment in our recent Letter \cite{prl}, extending the
calculations and presenting a more detailed account of the
theoretical methodology.

Our calculation is capable of quantitatively reproducing the unusual
excess DI yield of nitrogen molecules for a wide range of laser
intensities from $5\times 10^{13}$W/cm$^{2}$ to $1\times
10^{15}$W/cm$^{2}$. The significant influence of molecular alignment
on DI yield is virtually revealed: i) The ratio between double and
single ionization yield is less for perpendicular molecules than
that of parallel molecules; ii) This anisotropy becomes more
dramatic if the molecules are irradiated by a shorter laser pulse.
In particular, our model  provides an intuitive way of understanding
the complex dynamics involved in the molecular DI with back
analyzing  the classical trajectories. Their sub-cycle energy
evolution verify that collisions between electrons determines the
fate of molecular DI, therefore consolidate the classical
rescattering view of molecular DI. Statics based on plenty of
individual trajectories show that electrons are most likely to be
emitted at $30^{\circ }$ off laser peak,
indicating the accumulation of emitted electron pairs at $%
k_{1}^{||}=k_{2}^{||}=\pm 0.5 a.u. $ in the first and third
quadrants of parallel momentum plane $(k_{1}^{||}, k_{2}^{||})$,
which is consistent with the experimental results. Moreover, the
ejected electrons emerge in the same momentum hemisphere and
opposite momentum hemisphere alternately, depending on weather the
delayed time between closest collision and ionization is odd or even
half laser cycles.

Our  paper is organized as follows. In Sec.II we present our
semiclassical quasi-static model.  Calculations on double ionization
yield for nitrogen molecules  are presented and compared with
experimental data in Sec.III. In Sec.IV, we give an  analysis on the
classical trajectories and unveil sub-cycle dynamics of the
molecular double ionization
 Sec.V is our conclusions
and discussions.

\section{Model}

We consider a diatomic molecule with two valence electrons irradiated by a
laser field whose temporal and spatial distribution (see Fig.\ref{laser}(a))
is expressed as
\begin{equation}
\mathbf{\varepsilon}(t)=\varepsilon_{0}(R_{L},Z_{L})\sin^{2}(\frac{\pi t}{nT}%
)\cos(\omega t)\mathbf{e}_{z} .  \label{laser}
\end{equation}
In experiments, molecules are driven through the laser beam perpendicularly.
Thus the external field along the propagation direction of the laser beam is
approximately constant and in the lateral direction it can be treated as an
ideal Gaussian beam, i.e., $\varepsilon _{0}(R_{L},Z_{L})=\varepsilon
_{0}(R_{L})=\varepsilon _{0}\exp (-R_{L}^{2}/R_{0}^{2})$, where $\varepsilon
_{0}$ is the peak laser field, $R_{L}$ represents the position of the
molecules in the laser beam, and $R_{0}$ is the radius of the beam. The
maximum of $R_{L}$ is chosen to be three times or more of $R_{0}$ to ensure
the convergence of our results. A $sin^{2}$ enveloped laser with the full
width of $nT$ is used in our calculations, where T and $\omega $ are period
and angular frequency of the laser field respectively and n denotes the
number of optical cycles.
\begin{figure}[t]
\begin{center}
{\resizebox *{7.5cm}{11.0cm} {\includegraphics {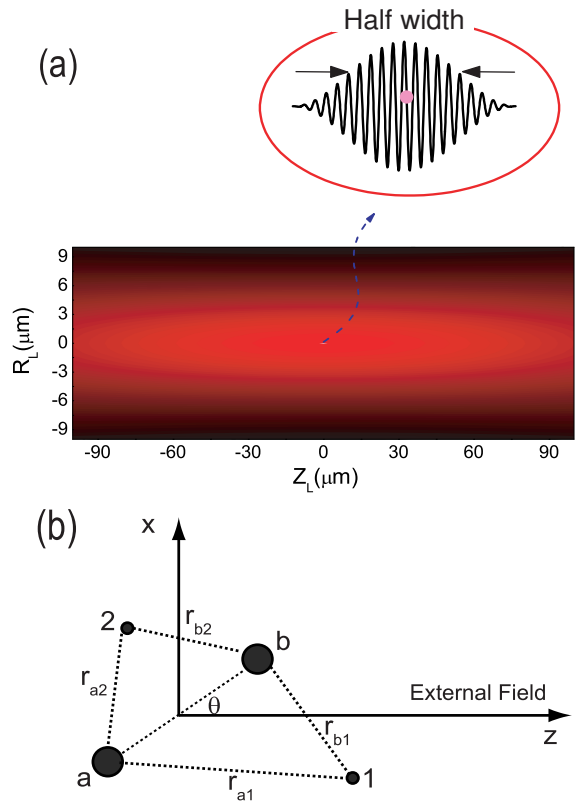}}}
\end{center}
\caption{(color online). (a) The spacial and temporal profile of a typical
laser beam, (b) The coordinate system adopted in our paper. }
\label{laser}
\end{figure}

In quantum mechanics, ionization processes are described by the
evolution of wave packets, however, in our model they are traced by
launching a set of trajectories with different initial parameters
instead (To some extent, the distribution of these parameters can be
regarded as an alternative emulation of quantum mechanics). For a
certain trajectory, the evolution of two electrons can be described
by classical Newton equations:
\begin{equation}
\frac{d^{2}\mathbf{r}_{i}}{dt^{2}}=\mathbf{\varepsilon }(t)-%
\bigtriangledown(V_{ne}^{i}+V_{ee}).  \label{Newton}
\end{equation}
Here $\varepsilon(t)$ is the external laser field discussed above. The index
i denotes the two different electrons. $V_{ne}^{i}$ and $V_{ee}$ are Coulomb
interaction between nuclei and electrons and between two electrons,
respectively.
\begin{align}
V_{ne}^{i} & =-\frac{1}{r_{ai}}-\frac{1}{r_{bi}},  \notag \\
V_{ee} & =\frac{1}{\left\vert \mathbf{r}_{1}-\mathbf{r}_{2}\right\vert },
\label{coulumb}
\end{align}
where $r_{ai}$ and $r_{bi}$ are distances between the ith electron and
nucleus a and b, respectively (as shown in Fig.\ref{laser}(b)).

To solve the Newton equations, we need the initial conditions of the
two electrons, including the initial time, initial position and
momentum. Their distribution are quite different for the tunneling
and over-the-barrier regime due to the different ionized mechanism
of the first electron.
\begin{figure}[t]
\begin{center}
\rotatebox{0}{\resizebox *{7.0cm}{6.7cm} {\includegraphics
{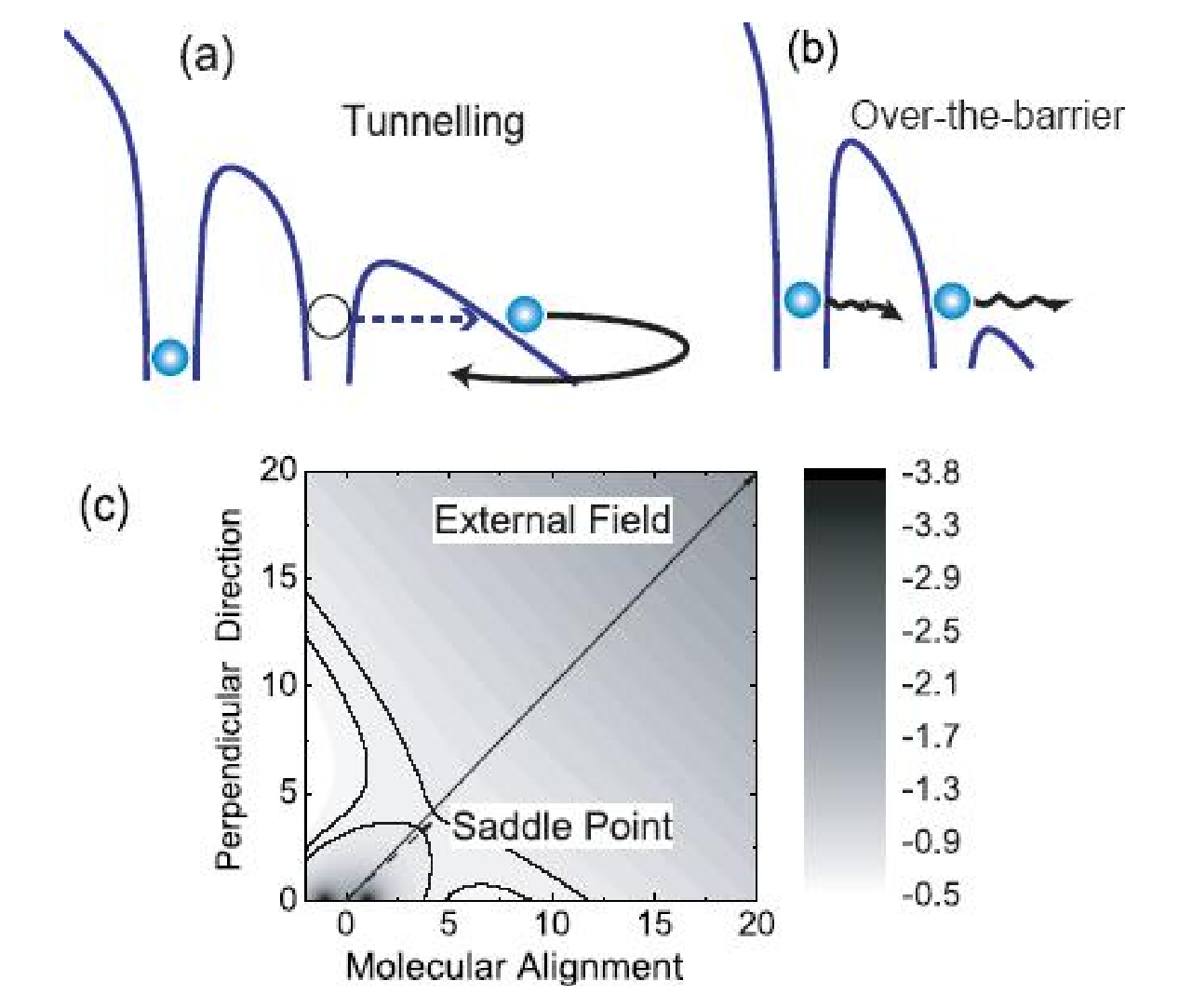}}}
\end{center}
\caption{(color online). (a) Tunneling ionization. (b) Over-the-barrier
ionization. (c) The contour plot of the combined Coulomb potential and
external laser field. It clearly shows that the saddle point locates
approximately along the direction of the external field.}
\label{firstion}
\end{figure}

\subsection{Tunneling regime}

In the long-wavelength limit, the laser field varies slowly in time
and can be regarded as a quasi-static field compared with valence
electron's circular motion around nuclei. Under this field, the
Coulomb potential between nuclei and electrons is significantly
distorted. When the instantaneous field (at time $t_0$) is
sufficient strong but still smaller than a threshold value (see
Fig.\ref{firstion}(a)), one electron is released at the outer edge
of the suppressed Coulomb potential through quantum tunneling with a
rate $\varpi(t_{0})$ given by molecular ADK formula\cite{adk}.

The electron tunnels out through a saddle point\cite{saddlepoint} directing
to a channel of the local minimum in the combined potential of the Coulomb
interaction and the external field (see, Fig.\ref{firstion}(c)). Because the
difference between the direction of the saddle point and the external field
is very small, we safely regard the external field direction ($z$ axis) as
the tunnelling direction. Thus, the initial position of the tunnelled
electron can be derived from following equation,
\begin{equation}
-\frac{1}{r_{a1}}-\frac{1}{r_{b1}}+\int\frac{\left\vert \Psi(\mathbf{r}%
^{^{\prime}})\right\vert ^{2}}{\left\vert \mathbf{r}_{1}\mathbf{-r}%
^{^{\prime }}\right\vert }d\mathbf{r}^{^{\prime}}+I_{p1}-z_{1}%
\varepsilon(t_{0})=0,  \label{position}
\end{equation}
with $x_{1}=y_{1}=0$. The wavefunction $\Psi$ is given by the linear
combination of the atomic orbital-molecular orbital (LCAO-MO) approximation.
Taking $N_{2}^{+}$ for example, we choose $\phi(r)=\frac{\lambda^{3/2}}{%
\sqrt{\pi}}e^{-\lambda r}$ as the trial function to construct the molecular
orbital $\Psi(r)=c[\phi(r_{a2})+\phi(r_{b2})]$, where c is the normalization
factor. The parameter $\lambda$, which equals to 1.54 for $N_{2}^{+}$, is
determined through variational approach. That is, calculate the variational
energy for the given wavefunction and assume it equals to the second
ionization energy of the molecule. The initial velocity of tunnelled
electron is set to be $(v_{\perp}\cos\varphi,v_{\perp}\sin\varphi,0)$, with $%
v_{\perp}$ having the same distribution as that in atomic case, i.e.,
\begin{equation}
w(v_{\perp})dv_{\perp}=\frac{2(2I_{p1})^{1/2}v_{\perp}}{\varepsilon(t_{0})}%
\exp(-\frac{v_{\perp}^{2}(2I_{p1})^{1/2}}{\varepsilon(t_{0})})dv_{\perp },
\label{velocity}
\end{equation}
where $\varphi$ is the polar angle of the transverse velocity uniformly
distributed in the interval $[0,2\pi]$.


In order to get the initial velocity distribution, we employ a technique
widely used in classical Monte Carlo(MC) simulation. Firstly, we generate
two random number $v_{\perp}^{test}$ and $w_{test}$ in the interval $%
[0,v_{\perp}^{\max}]$ and $[0,w_{\max}]$, respectively. If $%
w(v_{\perp}^{test})>w_{test}$, $v_{\perp}^{test}$ is kept as the initial
transverse velocity, otherwise it is rejected and the above procedure is
repeated.

For the bound electron, the initial position and momentum are
depicted by single-electron microcanonical distribution (SMD)
\cite{smd},
\begin{equation}
F(\mathbf{r}_{2},\mathbf{p}_{2})=k\delta\lbrack I_{p2}-\mathbf{p}%
_{2}^{2}/2-W(r_{a2},r_{b2})],  \label{micro1}
\end{equation}
where $k$ is the normalization factor, $I_{p2}$ denotes the ionization
energy of molecular ions such as $N_{2}^{+}$, and $%
W(r_{a2},r_{b2})=-1/r_{a2}-1/r_{b2}$ is the total interaction potential
between the bound electron and two nuclei.

\subsection{Over-the-barrier regime}

The above scheme is in the spirit of our quasi-static model for
atomic DI \cite{liuchenfu} and is applicable only when the
instantaneous laser field is weaker than the threshold value
\cite{cpl}. To give a complete description of the DI of molecular
system for the whole range of laser intensities (see Fig.\ref{37T}),
one needs to further extend
the above model to over-the-barrier regime (Fig.\ref{firstion}b, in such case Eq.(%
\ref{position}) has no real roots). This is done by constructing the
initial conditions with double-electron microcanonical distribution
(DMD) \cite{dmd}, i.e.,
\begin{align}
F(\mathbf{r}_{1},,\mathbf{r}_{2},\mathbf{p}_{1},\mathbf{p}_{2}) & =\frac {1}{%
2}[f_{\alpha}(\mathbf{r}_{1}\mathbf{,p}_{1})f_{\beta}(\mathbf{r}_{2}\mathbf{%
,p}_{2})  \notag \\
& +f_{\beta}(\mathbf{r}_{1}\mathbf{,p}_{1})f_{\alpha}(\mathbf{r}_{2}\mathbf{%
,p}_{2})],  \label{micro2}
\end{align}
with
\begin{equation}
f_{\alpha,\beta}(\mathbf{r,p})=k\delta\lbrack I_{p1}-\frac{\mathbf{p}^{2}}{2}%
-W(r_{a},r_{b})-V_{\alpha,\beta}(\mathbf{r)}],  \label{micro3}
\end{equation}
where $V_{\alpha ,\beta }(\mathbf{r)}$ represents the mean interaction
between two electrons, $V_{\alpha ,\beta }(\mathbf{r)=}\frac{1}{r_{b,a}}%
[1-(1+\kappa r_{b,a})e^{-2\kappa r_{b,a}}]. $ $\kappa $ can be
obtained by a variational calculation of the ionization energy of
molecules ($\kappa=1.14$ for $N_{2}$). Details can be found in
Ref.\cite{dmd}.

We would like to give some remarks here. (i) In our calculations,
the trajectories obtained from DMD are weighed by $\varpi(t_{0})$
\cite{adk}.
(ii) Part of the electrons obtained from DMD could "self-ionize"
even without the presence of external field. To avoid such a
unphysical self-ionization, we evolute the electrons for several
optical cycles freely and abandon those samples  whose energy are
greater than zero during the free evolution process.
\begin{figure}[t]
\begin{center}
\rotatebox{0}{\resizebox *{8.0cm}{6.0cm} {\includegraphics
{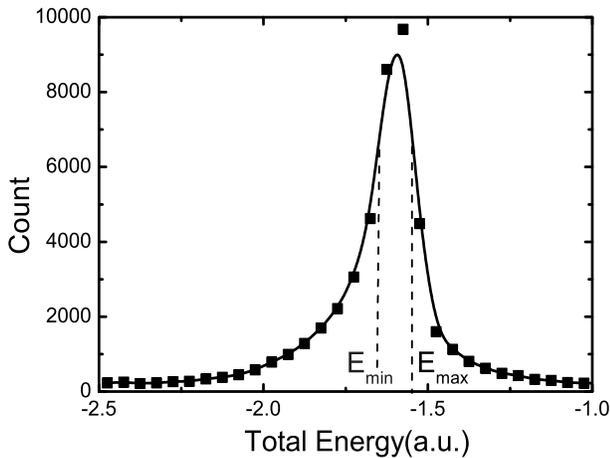}}}
\end{center}
\caption{Total energy distribution of the two electrons. $E_{min}$ and $%
E_{max}$ are used to cut off the long tail. Only the electrons with
a total energy between $E_{min}$ and $E_{max}$ is adopted in our
calculation.} \label{energy}
\end{figure}
(iii) The initial total energy distribution of the two electrons is
plotted in Fig.\ref{energy}, where we see a long tail on both sides.
They are cut off by introducing two parameters $E_{min}$ and
$E_{max}$, which satisfy
\begin{equation}
{\displaystyle}\frac{{\int\nolimits_{E_{\min }}^{E_{\max }}}E\rho (E)dE}{{%
\int\nolimits_{E_{\min }}^{E_{\max }}}\rho (E)dE}=E_{M},  \label{energy}
\end{equation}%
where $E_{M}\approx I_{p1}+I_{p2}$ is the most probable energy and $\rho (E)$
is the state density around $E$.

With the above initial conditions, the Newtonian equations are
solved using the 4-5th step-adaptive Runge-Kutta algorithm and DI
events are identified by energy criterion. In our calculations, more
than $10^5$ weighted (i.e. by rate $\varpi(t_{0})$) classical
trajectories of electron pair are traced and a few thousands or more
of DI events are collected for statistics. Convergency of the
numerical results is further tested by increasing the number of
launched trajectories twice.

\section{ Calculation on the DI of N$_{2}$: comparison with experiments}
In this section, we apply  our theory to the nitrogen molecules and
compare our model calculation  with experimental data. The
parameters are chosen to match the experiments, \textit{e.g.} the
internuclear separation is
2.079a.u., the first and second ionization energy are $I_{p1}=0.5728a.u.$, $%
I_{p2}=0.9989a.u. $, respectively. The laser frequency is $\omega
=0.05695a.u.$ corresponding to a wavelength of 800nm. Based on our
model, we successfully reproduce the plateau representing the
excessive DI yields due to the correlation between electrons , and
reveal  a slight shallow dip at intensity of $0.4\times
10^{15}W/cm^2$ indicating a transition point below which  the
electron correlation plays role. In addition, our model calculations
confirm the dramatic dependence of DI yield on molecular alignment.
\begin{figure}[t]
\begin{center}
\rotatebox{0}{\resizebox *{7.5cm}{6.5cm} {\includegraphics
{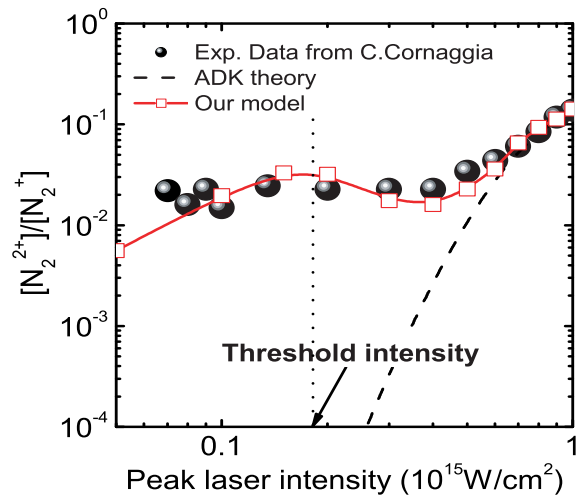}}}
\end{center}
\caption{Comparison between DI data\protect\cite{cornaggia} and
theory for nitrogen molecule. 0.185 PW/cm$^2$ is the threshold
intensity separates the tunneling regime and over-the-barrier regime
as schematically plotted in Fig.\protect\ref{firstion}. To our
knowledge, the results from our theoretical model are the first to
be in good agreement with experimental data for a wide range of
laser intensities from tunneling regime to over-the-barrier regime.
} \label{37T}
\end{figure}
In Fig.\ref{37T}, the ratio between double and single ionization
yield is
plotted with respect to the peak laser intensities from $5\times 10^{13}$W/cm%
$^{2}$ to $1\times 10^{15}$W/cm$^{2}$. The number of optical cycle
is chosen as 37 to match the experiments of Cornaggia
\cite{cornaggia}. Our numerical results show good agreement with the
experimental data over the whole range, thus confirm the validity of
our model. The curve can be clearly divided into three parts: the low intensity regime (%
$5\times10^{13}$W/cm$^{2}$ to $1.85\times10^{14}$W/cm$^{2}$), where
tunneling ionization plays an important role; the very high intensity regime ($%
5\times10^{14}$W/cm$^{2}$ to $1\times10^{15}$W/cm$^{2}$) which can
be also well described by ADK theory, indicating that the electrons
are pulled out sequentially. In particular, there exists a plateau
(moderate intensity) regime with a shallow dip, which can be seen
from the corresponding experiments in Ref.%
\cite{cornaggia} Fig.10. Back analysis about classical trajectories
indicates that the underlying dominant mechanism in these three
regimes are widely divergent as will be shown latter.

\begin{figure}[t]
\begin{center}
\rotatebox{0}{\resizebox *{7.5cm}{3.5cm} {\includegraphics
{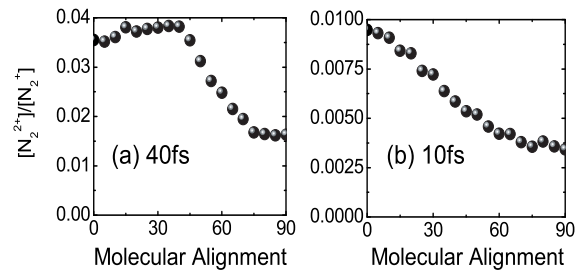}}}
\end{center}
\caption{The molecular alignment dependence of DI ratios for laser
intensity of 0.15PW/cm$^{2}$.} \label{angle}
\end{figure}

In our producing data in Fig.\ref{37T} the molecular alignment is
set to be random and our result is obtained by averaging over
different orientations. However, the inherent nuclear degrees of
freedom of molecules do manifest themselves as the significant
alignment effect in our model. On the other hand, recent progress in
experimental technique has make it possible to control molecular
alignment by applying a weak pre-pulse \cite{zeidler}. With these
considerations and further application of our model, we calculate
the ratios between double and single ionization according to
different angles between molecular alignment and laser polarization.
Main results are presented in Fig.\ref{angle}. It clearly shows two
tendencies, i) The ratio between double and single ionization yield
is less for perpendicular molecules than that of parallel molecules;
ii) This anisotropy becomes more dramatic when the molecules are
irradiated by a shorter laser pulse.

Compared with shorter pulse case, Fig.\ref{angle}(a) shows the
ratios between double and single ionization keeps almost a constant until $%
35^{\circ }$ and then decreases rapidly. At $90^{\circ }$ (i.e.,
perpendicular case), the ratio become half of that in parallel case.
This result is qualitatively in agreement with experimental data
\cite{zeidler}. Quantitatively, our factor 2 is larger than the
observed ratio of 1.1 between parallel and perpendicular case
\cite{zeidler}. This discrepancy comes from the unperfect control of
molecular alignment in practical experiments due to the rotation of
molecules and many other uncertain factors. Based on our theoretical
results, a simple estimation will conclude that only about $15\%$ of
the molecules are "switched" to the appointed direction
additionally. Further explorations show that molecular alignment
also significantly affects the correlated momentum distribution of
emitted electrons. Details will be presented elsewhere \cite{li}.

\section{Sub-cycle dynamics of molecular DI}
With our model we are capable  to extract vital information by
tracing back the history of individual DI trajectories\cite{haan}.
Such classical trajectory perspective provides us an intuitive way
towards understanding the complex dynamics involved in molecular DI.

\subsection{Typical trajectories responsible for emitting electron pairs}

\begin{figure}[t]
\begin{center}
\rotatebox{0}{\resizebox *{8.0cm}{10.5cm} {\includegraphics
{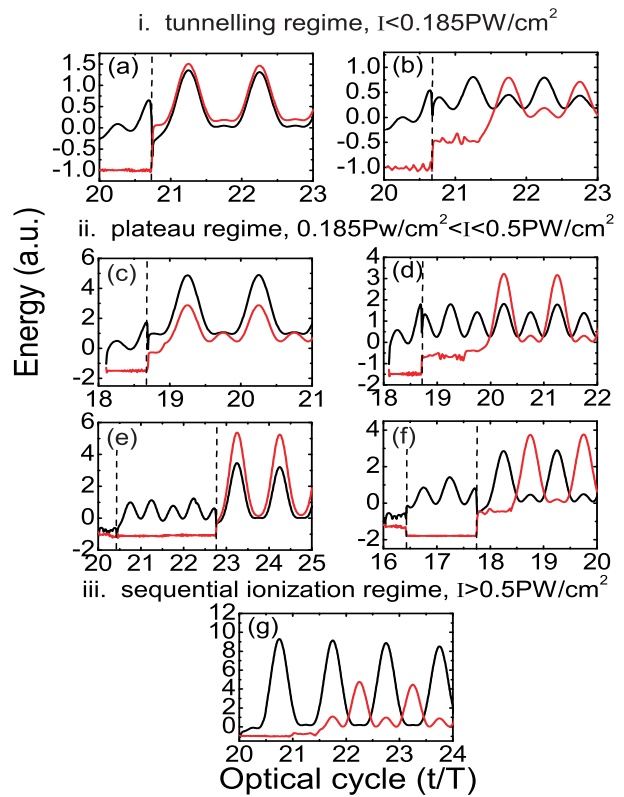}}}
\end{center}
\caption{(color online). Typical energy evolution of the electron pair in
three different regimes: (i) tunneling regime, (ii) plateau regime and (iii)
sequential ionization regime. Vertical dashed lines indicate the moment when
significant collision between two electrons emerge.}
\label{evl}
\end{figure}

In Fig.\ref{37T}, the vertical dot line indicates the threshold
value of 0.185 PW/cm$^2$, which separates the double ionization data
into two parts, the tunneling regime and over-the-barrier regime.
When peak laser intensity is below this value, there exist two
dominant processes responsible for emitting both electrons, namely,
collision-ionization(CI) and collision-excitation-ionization(CEI),
as shown in Fig.\ref{evl}(a)(b), respectively. The unique feature of
these two mechanism is that the first electron tunnels out through
the potential barrier and escapes instead of being pulled out
directly by the laser field. This tunneling mechanism can't be
included by any classical theory, thus we have not choice but place
an electron outside the barrier directly. As a result, we usually
notice that the tunneled electron has a higher initial energy as
shown in Fig.\ref{evl}(a) and (b); while in other trajectories, the
two electrons almost have the same initial energy, a reflection of
the identity. For CI, the tunnelled electron is driven back by the
oscillating laser field to collide with the bounded electron near
its parent ions causing an instant ($\sim$ attosecond) ionization.
The condition for occurrence of CI is usually hard to reach. 
It requires the kinetic energy of the returned electron exceeds the
ionization energy of the bound electron and they should move close
to each other to share the excess kinetic energy. However, as
pointed out by other numerical studies, the returning trajectories
can be assumed to be 
a freely spreading Gaussian wave packet whose return width is about
30a.u.\cite{returnwidth}. So most of the cases, the returned
electron just slides apart from the bound electron and only
contribute a little proportion of its kinetic energy to the inner
one, in this case CEI occurs instead. For CEI, the inner electron is
firstly excited through collision with the returned electron, then
undergoes a time-delayed ($\sim$ a few optical periods) field
ionization of the excited state. During the time interval, the
returned and escaped again electron oscillates in the laser field
like a near-free electron while the excited but still bounded
electron move around the cores until accumulate enough energy to
cause DI.

When the instantaneous laser field is above the threshold value,
over-barrier ionization emerges. In this regime we observe more complicated
trajectories for DI processes. Except for CI (Fig.\ref{evl}%
(c)) and CEI (Fig.\ref{evl}(d)) trajectories similar to tunneling
cases, there are multiple-collision trajectories as shown in
Fig.\ref{evl}(e),(f) as well as collisionless trajectory of
Fig.\ref{evl}(g). In Fig.\ref{evl}(e) and (f), two valence electrons
entangle each other initially , experience a multiple-collision and
then emitted. These two kinds of trajectories exits around the
threshold intensity where the over-barrier ionization start to
emerge but the external field is not strong enough to pull out the
electron immediately without the assistant of Coulomb repulsion
between two electrons. The four types of trajectories indicated by
Fig.\ref{evl}(c-f) represent the dominant processes of DI in the
plateau regime from 0.185PW/cm$^2$ to 0.5PW/cm$^2$, which are much
more complicated than that of tunneling regime, but still
accompanied by once or multiple times of collisions between two
electrons \cite{paulus}. However, above 0.5PW/cm$^2$, the double
ionization is dominated by a collisionless sequential ionization
whose typical trajectory is represented by Fig.\ref{evl}(g). In this
regime, our numerical results tend to ADK theory.

\subsection{Time delay effect on correlated momentum}
A key feature distinguishing classical (or semi-classical) model
from quantum-mechanical treatment is the most conspicuous and least
ambiguous implication of three time scale during the history of
individual DI trajectories \cite{timedelay}, i.e., the time of
single ionization, the time of closest collision and the time of DI.
The time of single ionization is adopted as the start time of our
simulation, and determines the weight of each trajectory. After
single ionization, the electron travel much of the time in the
intense laser field like a classical object until recollide with the
bound electron, thus lead to DI eventually. Usually, there is a time
lag between the time of DI and closest collision. Such time delay
determines the fate of two electrons, and is also firmly related to
the final parallel momentum.

Like the atomic cases, a simplified model\cite{liuchenfu} neglecting
most of the unimportant factors, such as the spacial distribution of
the laser beam and temporal envelop of the laser pulse, would be
more convenient while still give clear interpretation to most of the
phenomena. So in this section, we assume the molecules are exposed
to a cosine external field and the first electrons tunnelled out in
the first half optical cycle. The electric field is switched off
using a $cos^{2}$ envelope during two optical cycles at last. As we
would discuss later, many important information can be extracted
from the distribution of laser phase at the moment of closest
collision and ionization \cite{feuerstein, weckenbtock}. We choose
three typical laser intensities, i.e. $0.12$PW/cm$^{2}$,
$0.4$PW/cm$^{2}$ and $1$PW/cm$^{2}$, representing the tunneling
regime, plateau regime and sequential ionization regime, to analyze
in detail.
\begin{figure}[t]
\begin{center}
\rotatebox{0}{\resizebox *{8.0cm}{3.5cm} {\includegraphics
{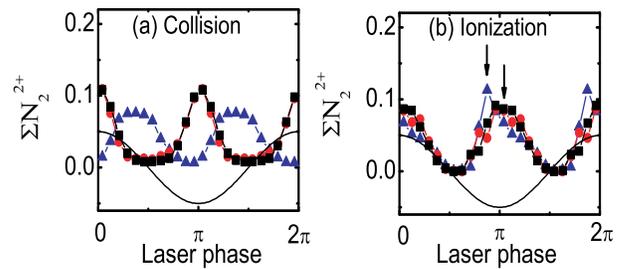}}}
\end{center}
\caption{(color online). DI yield vs laser phase when (a) the two
electrons become closest; (b) both the electrons are ionized, at
different laser intensity $0.12$PW/cm$^{2}$(triangle), $0.4$PW/cm$^{2}$%
(circle) and $1$PW/cm$^{2}$(square), respectively. The solid lines represent
the laser field.}
\label{timeall}
\end{figure}

Fig.\ref{timeall}(a) shows the diagram of DI yield versus laser
phase at the moment of closest collision. In the tunneling regime
(i.e., $0.12$PW/cm$^{2}$), we note that the collision can occur
throughout most of the laser cycle and the peak emerges slightly
before the zeroes of
the laser field, consistent with the prediction of simple-man model\cite%
{corkum} and recent results from purely classical
calculation\cite{haan}. However, for other two cases, the collision
between two correlated electron turns to occur mainly at peak laser
field. This is because ionization mechanism changes at the
transition to over-the-barrier regime, where both electrons rotate
around nuclei initially and they usually become closest before one
of them is driven away by the external field (laser peak) rather
than during the recollision process (zeroes of the laser field).

Fig.\ref{timeall}(b) confirms that most DI occurs around the maximum
of laser field for both tunneling regime and over-the-barrier
regime. More interestingly, compared to two other cases we observe a
peak shift of $\sim 30^o$ off the field maximum for the tunneling
case. It is due to the larger fraction of CI trajectories in this
regime. With assuming that the colliding electrons leave the atom
with no significant energy and electron-electron momentum exchange
in final state is negligible (these assumptions have been checked by
directly tracing trajectories), the parallel momentum $k_{1,2}^{||}$
of each electron results exclusively from the acceleration in the
optical field: $k_{1,2}^{||}=\pm 2\sqrt U_p \sin\omega t_{ion}$
\cite{weckenbtock}. The above shifted peak indicates the
accumulation of the emitted electrons at $k_{1}^{||}=k_{2}^{||}=\pm
0.5 a.u.$ in the first and third quadrants of parallel momentum
plane $(k_{1}^{||},
k_{2}^{||})$. It is consistent with the experimental data of Ref.\cite%
{zeidler} (see their Fig.2).

\begin{figure}[t]
\begin{center}
\rotatebox{0}{\resizebox *{7.5cm}{6.5cm} {\includegraphics
{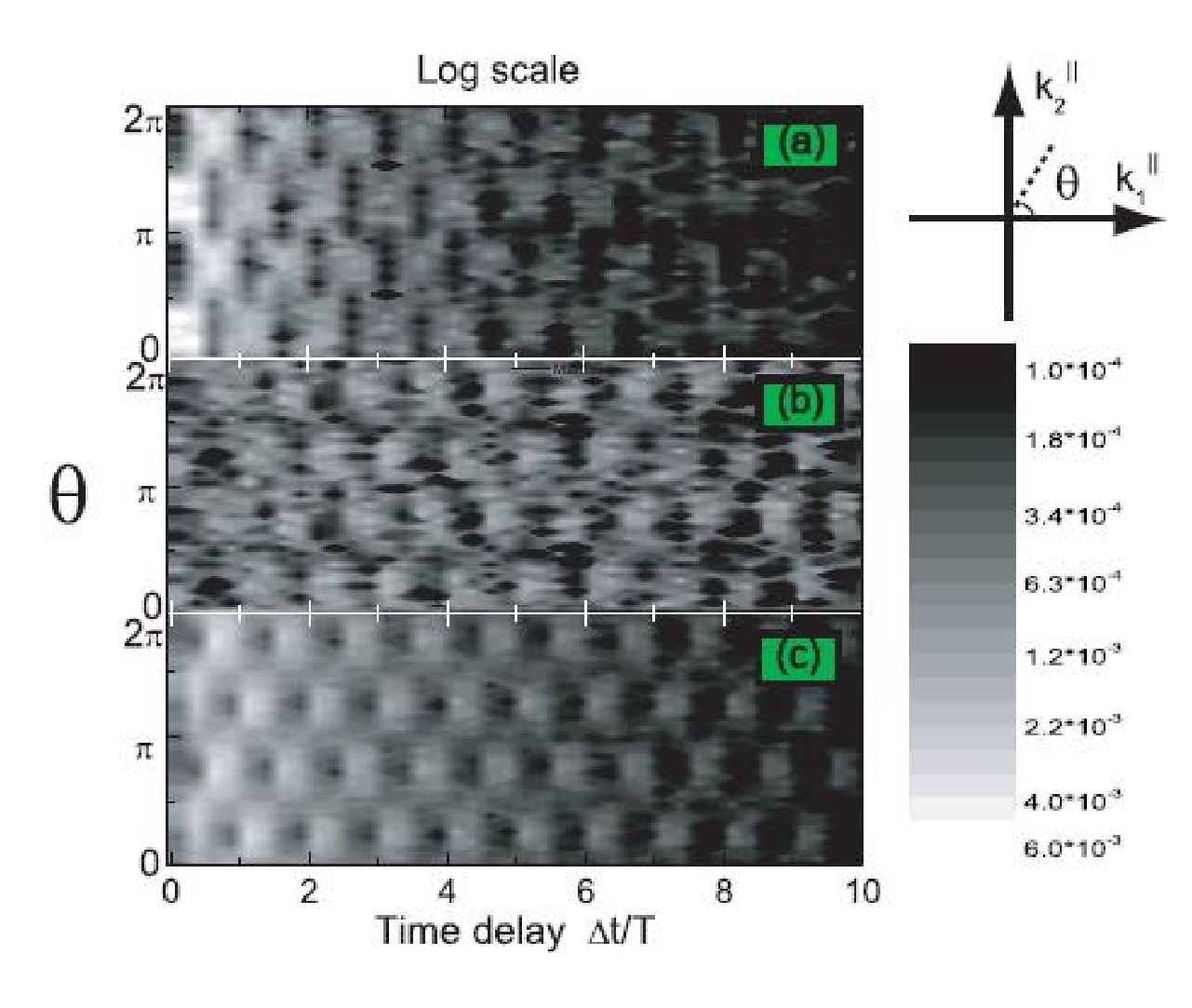}}}
\end{center}
\caption{(color online). The relationship between the correlated momentum
and the delay time at (a)$0.12$PW/cm$^{2}$ (b)$0.4$PW/cm$^{2}$ (c)$1.0$PW/cm$%
^{2}$ }
\label{momentum}
\end{figure}
Usually, there is a time delay between closest collision and ionization. Fig.%
\ref{momentum} shows the profound effects of time delay on correlated
momentum, where the phase angle of momentum vector $(k_{1}^{||},k_{2}^{||})$
is introduced to describe the correlation between two electrons' momentum: $%
[0,\pi/2]$ and $[\pi,3\pi/2]$ correspond to a same-hemisphere
emission while $[\pi/2,\pi]$ and $[3\pi/2,2\pi]$ correspond to a
opposite-hemisphere emission. In all three cases we observe a
long-tail up to several optical cycles. For the sequential
ionization of 1 PW/cm$^2$, it means after one electron is deprived
from nuclei by laser fields the other electron is slowly (i.e.
waiting for up to a few optical cycles) ionized. In the tunneling
regime, the long-tail indicates that CEI mechanism is very
pronounced for the molecular DI (contribute to 80\% of total DI
yield). This observation is different from purely classical
simulation \cite{haan}, where CI effect is believed to be
overestimated. Our
results, however, are consistent with experiments of Ar atom \cite%
{feuerstein}, where the ionization energy and laser field parameters are
close to our case. The reason is stated as follows. For the intensity of $%
0.12$PW/cm$^{2}$, the maximal kinetic energy of the returned electron is $%
3.17U_{p}=0.85a.u.$, still smaller than the ionization energy of
$N_{2}^+$. Even with the assistance of the Coulomb focusing
\cite{corkum,coulumb}, it is still not easy for the returned
electrons to induce too many CI events.

Furthermore, such time delay might provide more physics beyond simple
rescattering scenario. Recently a statistical thermalization model has been
proposed for the nonsequential multiple ionization of atoms in the tunneling
regime \cite{attothermal}. This model shows that sharing of excess energy
between the tunnelled electron and the bound electrons takes some time,
resulting in a time delay on attosecond time scale between recollision and
ionization. Our simulation upholds this picture of attosecond electron
thermalization: on upper panel of Fig. \ref{momentum}, two bright spots are
observed at a similar time delay on subfemtosecond time scale for CI
trajectory.


\begin{figure}[t]
\begin{center}
\rotatebox{0}{\resizebox *{6.5cm}{6.5cm} {\includegraphics
{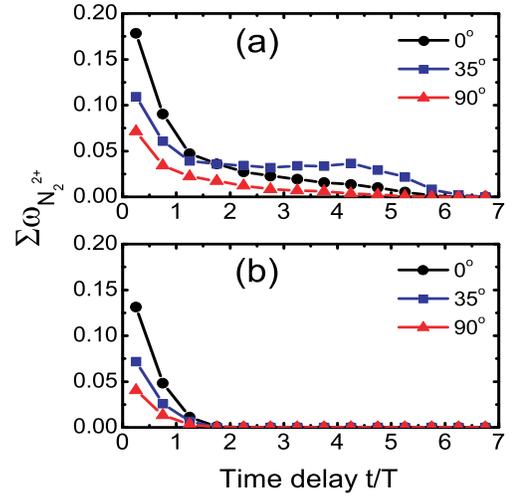}}}
\end{center}
\caption{(color online). DI yield versus the delayed time at laser
intensity of 0.15PW/cm$^{2}$. The total optical cycles is (a)7,
(b)3, respectively. } \label{why35}
\end{figure}

The regular pattern in Fig.\ref{momentum}(a),(c) exhibit that the ejection
of electrons in the same-hemisphere and opposite-hemisphere emerge
alternately with respect to the delayed time. For a time delay of odd half
laser cycles, two electrons are emitted in the same direction, for a time
delay of even half laser cycles, two electrons are emitted back-to-back.
Moreover, in the tunneling regime the pattern in Fig.\ref%
{momentum}(a) shows two notably bright spot in the first and third quadrants
when the delayed time is less than 0.5T, a phenomenon directly due to the CI
trajectories. On the other hand, the irregular pattern emerges in Fig.\ref%
{momentum}(b) for DI in the plateau regime as the signature of
complicated multiple-collision trajectories. It implies that the
trajectories of two electrons entangle with each other before DI
ionization occurs and the electrons' motion might be chaotic
\cite{liuhu}.

Integrating over angle in Fig.\ref{momentum} gives total DI yield vs
the delayed time as illustrated in Fig.\ref{why35}, where a laser
field with seven and three optical cycles is used in (a) and (b),
respectively. In each case, we also choose three alignment angles, i.e. $%
0^{\circ }$, $35^{\circ }$ and $90^{\circ }$ for comparison. As
known to all, a DI event with delayed time less than 0.5T represents
CI trajectory while others are all CEI trajectories. In
Fig.\ref{why35}(a), both CI and CEI are suppressed for perpendicular
molecules compared with parallel ones. However, for the case of
$35^{\circ }$, CEI play an more important role than other two cases.
As a result, the total number of CI and
CEI events stays almost unchanged and forms the quasi-plateau between [$0^{\circ }$,%
$35^{\circ }$] in Fig.\ref{angle}(a). This plateau can eliminated by using
shorter laser pulses because the excessive CEI can be effectively suppressed
as demonstrated in Fig.\ref{angle}(b).

\section{Conclusions}

In summary, we have developed a semiclassical quasi-static model
with including classical rescattering and quantum tunneling effects
and  achieved insight into the dynamics of two highly-correlated
valence electrons under the combined influence of a two-center
Coulomb potential and an intense laser field. Our model can be used
to compared with experimental data of molecular DI quantitatively
under the relevant experimental conditions, i.e., highly
nonperturbative fields with femtosecond or shorter time resolution.
We reproduce the experimental data of the  plateau structure of the
DI, uneven DI yield for parallel or  perpendicular aligned
molecules, and the accumulation of the emitted electrons in the
first and third quadrants of parallel momentum plane. In addition,
our model calculation reveals some novel features, such as a shallow
dip in the plateau regime, enhancement of anisotropy when
illuminated by shorter laser pulses and the  regular patterns of
correlated momentum with respect to the time delay. We hope our
theory will stimulate the experimental works in the direction.


\section{Acknowledgments}

This work is supported by NNSF of China No.10574019, CAEP Foundation
2006Z0202, and 973 research Project No. 2006CB806000. We thank J. H.
Eberly for stimulating discussions. In particular, we  are indebted
to X. Liu for very useful discussions and suggestions.


\begin{thebibliography}{99}

\bibitem{review} M. Protopapas, C. H. Keitel, and P. L. Knight, Rep. Prog.
Phys. \textbf{60}, 389 (1997).

\bibitem{book} \emph{Atoms in Intense Laser Fields}, edited by M. Gavrila
(Academic, New York, 1992).

\bibitem{nature} B. Walker, B. Sheehy, L. F. DiMauro \emph{et al.}, Phys. Rev.
Lett. \textbf{73}, 1227 (1994); Th. Weber \textit{et al.}, Nature
\textbf{405}, 658 (2000); X. Liu, H. Rottke, E. Eremina \emph{et
al.}, Phys. Rev. Lett. \textbf{93}, 263001 (2004), and references
therein.

\bibitem{landau} L. D. Landau and E. M. Lifshitz, \emph{Quantum Mechanics}
(Pergamon Press, New York, 1977).



\bibitem{shakeoff} D. N. Fittinghoff, P. R. Bolton, B. Chang and K. C.
Kulander, Phys. Rev. Lett. \textbf{69}, 2642 (1992).

\bibitem{corkum} P. B. Corkum, Phys. Rev. Lett. \textbf{71} , 1994 (1993).

\bibitem{coulumb} T. Brabec, M. Y. Ivanov, and P. B. Corkum,
Phys. Rev. A \textbf{54}, R2551 (1996).

\bibitem{rescatter} Th. Weber \emph{et al.}, Phys. Rev. Lett. \textbf{84},
443 (2000); R. Moshammer \emph{et al.}, Phys. Rev. Lett. \textbf{84}, 447
(2000); A. Becker and F. H. M. Faisal, Phys. Rev. Lett. \textbf{84}, 3546
(2000); M. Lein \emph{et al.}, Phys. Rev. Lett. \textbf{85}, 4707 (2001).

\bibitem{guo} C. Guo, M. Li, J. P. Nibarger, and G. N. Gibson Phys. Rev. A
\textbf{58}, R4271 (1998).

\bibitem{cornaggia} C. Cornaggia and Ph. Hering, Phys. Rev. A \textbf{62},
023403 (2000).

\bibitem{soft} J. Muth-B\"ohm, A. Becker, and F. H. M. Faisal, Phys. Rev.
Lett. \textbf{85}, 2280 (2000).

\bibitem{dissociate} T. K. Kjeldsen and L. B. Madsen, J. Phys. B: At. Mol.
Opt. Phys. \textbf{37}, 2033 (2004).

\bibitem{alnaser} A. S. Alnaser, S. Voss, X. M. Tong \emph{et al.}, Phys.
Rev. Lett. \textbf{93}, 113003 (2004)

\bibitem{eremina} E. Eremina, X. Liu, H. Rottke \emph{et al.}, Phys. Rev.
Lett. \textbf{92}, 173001 (2004)

\bibitem{zeidler} D. Zeidler, A. Staudte, A. B. Bardon, D. M. Villeneuve, R.
D\"o rner, and P. B. Corkum, Phys. Rev. Lett. \textbf{95}, 203003 (2005).

\bibitem{taylor} J. S. Parker \emph{et al.}, J. Phys. B \textbf{33}, L691
(2000).

\bibitem{pegarkov} A. I. Pegarkov, E. Charron and A. Suzor-Weiner, J. Phys. B
\textbf{32}, L363(1999).

\bibitem{beck} A. Becker and F. H. M. Faisal, J. Phys. B \textbf{38}, R1
(2005).

\bibitem{saddlepoint} Jakub S. Prauzner-Bechcicki, Krzysztof Sacha, Bruno
Eckhardt, and Jakub Zakrzewski, Phys. Rev. A \textbf{71} , 033407 (2005).

\bibitem{prl} J. Liu, D. F. Ye, J. Chen, and X. Liu, Phys. Rev. Lett.
\textbf{99}, 013003 (2007).

\bibitem{adk} The atomic ADK theory has been extended to diatomic
molecules; see, for example, X. M. Tong \emph{et al.}, Phys. Rev. A \textbf{%
66} , 033402 (2002) and I. V. Litvinyuk \emph{et al.}, Phys. Rev.
Lett. \textbf{90}, 233003 (2003). An explicit analytic expression
has also been derived in \cite{li}. However, we found that the employment of
atomic ADK formula instead of the complicated molecular ADK formula does not lead
to significant discrepancy in calculating the ratios between double and single
ionization. So, for simplicity, we adopt $%
\varpi(t_{0})=\frac{4(2I_{p1})^{2}}{\varepsilon(t_{0})}\exp(-\frac{%
2(2\left\vert I_{p1}\right\vert )^{3/2}}{3\varepsilon(t_{0})})$ in
our calculations.

\bibitem{li} Y. Li, J. Chen, S. P. Yang, and J. Liu, 'Correlated momentum
distribution of doule-ionized molecules', to appear in Phys. Rev. A
(2007).

\bibitem{smd} R. Abrines and LC. Percival, Proc. Phys. Soc. London \textbf{88%
} , 861 (1966); J. G. Leopold and I. C. Percival, J. Phys. B \textbf{12} ,
709 (1979).

\bibitem{liuchenfu} Li-Bin Fu, Jie Liu, Jing Chen, and Shi-Gang Chen Phys.
Rev. A \textbf{63}, 043416 (2001); J. Chen, J. Liu, L. B. Fu, and W.
M. Zheng Phys. Rev. A \textbf{63}, 011404 (2001); Li-Bin Fu, Jie
Liu, and Shi-Gang Chen Phys. Rev. A \textbf{65}, 021406 (2002); J.
Chen, J. Liu, and W. M. Zheng Phys. Rev. A \textbf{66}, 043410
(2002).

\bibitem{cpl} Jie Liu and Jing Chen, Chin. Phys. Lett. \textbf{23} 91
(2006).

\bibitem{dmd} L. Meng, C. O. Reinhold and R. E. Olson, Phys. Rev. A \textbf{%
40} , 3637 (1989).




\bibitem{haan} S. L. Haan, L. Breen, A. Karim, and J. H. Eberly, Phys. Rev.
Lett. \textbf{97}, 103008 (2006), and references therein.

\bibitem{returnwidth} B. Walker \textit{et al.}, Phys.
Rev. Lett. \textbf{73} , 1227 (1994); B. Walker \textit{et al.},
Phys. Rev. Lett. \textbf{77} , 5031 (1996).

\bibitem{paulus} G. G. Paulus, W. Becker, W. Nicklich and H. Walther, J.
Phys. B \textbf{27}, L703 (1994).

\bibitem{timedelay} J. S. Parker, B. J. S. Doherty, K. J. Meharg, and K. T. Taylor,
J. Phys. B \textbf{36}, L393 (2003).

\bibitem{feuerstein} B. Feuerstein, R. Moshammer, D. Fischer \textit{et al.}, Phys.
Rev. Lett. \textbf{87} , 043003 (2001).

\bibitem{weckenbtock} M. Weckenbrock, D. Zeidler, A. Staudte \textit{et al.},
Phys. Rev. Lett. \textbf{92}, 213002 (2004).

\bibitem{attothermal} X. Liu, C. Figueira de Morisson Faria, W. Becker and
P. B. Corkum , J. Phys. B \textbf{39}, L305 (2006).

\bibitem{liuhu} B. Hu, J. Liu and S.G Chen, Phys. Lett. A \textbf{236}, 533
(1997)
\end{thebibliography}
\end{document}